\documentclass[12pt,preprint]{aastex}
\usepackage{epsfig}
\begin{document}

\title{Probing the Inflow/Out-flow and Accretion Disk of Cyg X-1 in the High State
      with HETG/Chandra}
\author{Y.X. Feng$^{1,2}$, A.F. Tennant$^3$ and S.N. Zhang$^{1,2}$}

\affil{$^1$Department of Physics, University of Alabama in Huntsville, Huntsville,
AL. 35899\\
$^2$ National Space Science and Technology Center, NASA Marshall Space Flight Center,
SD50, 320 Sparkman Dr., Huntsville, AL 35805\\
$^3$ NASA Marshall Space Flight Center, SD50, Huntsville, AL 35812\\
e-mail:fengy@email.uah.edu, Allyn.Tennant@msfc.nasa.gov, zhangsn@email.uah.edu}

\begin{abstract}
Cyg X-1 was observed in the high state at the conjunction orbital phase (0)
with HETG/Chandra.
Strong and asymmetric absorption lines of highly ionized
species were detected, such as Fe XXV, Fe XXIV, Fe XXIII, Si XIV,
S XVI, Ne X, and etc.
In the high state the profile of the absorption lines are composed of an
extended red wing and a less extended blue wing.
The red wings of higher ionized species are more extended than that of
lower ionized species.
The detection of these lines provides a way to probe the properties of the flow around
the companion and the black hole in Cyg X-1 during the high state.
A broad emission feature around 6.5 keV was  significantly detected
from the both spectra of HETG/Chandra and PCA/RXTE.
This feature appears to be symmetric and can be fitted with a Gaussian function
rather than the Laor disk line model of fluorescent Fe K$_\alpha$ line from an accretion disk.
The implications of these results on the structure of the accretion flow of Cyg X-1
in the high state are discussed.
\end{abstract}


\section{Introduction}

Cyg X-1 is a black hole candidate which consists a compact object orbiting
a $\sim$17 M$_{\odot}$ O9.7 I supergiant with a period of 5.6 day (Webster
and Murdin 1972; Bolton 1972).
The dynamically determined mass of the compact object which exceeds significantly the neutron
star upper limit provides evidence of a black hole in Cyg X-1.
The companion star almost fills its Roche lobe in Cyg X-1 (Gies and Bolton 1986).
Therefore, the configuration of material transferring from the companion to the black hole
may be intermediate between Roche-lobe overflow and stellar wind accretion.
The X-ray emission of Cyg X-1 is believed to be powered by the accreted
stellar wind material from the companion.
The X-ray spectrum is composed of soft and hard spectral components
(see, e.g., Liang \& Nolan 1984, Tanaka \& Lewin 1995). Various spectral shapes
 have been found and classified as the so called, low, intermediate and high spectral states
(see, e.g., Oda 1977; Liang \& Nolan 1984, and Tanaka \& Lewin 1995; Belloni et al. 1996).
 Primarily, this classification is based on the power-law index obtained by fitting the soft
(2 - 10 keV) X-ray spectrum with a single power-law. In the low state, the power-law index
is about 1.6. In the intermediate state, the power-law index is around 2.0 and it is about 2.5
in the high state ( see, e.g., Oda 1977; Liang \& Nolan 1984, and Tanaka \& Lewin 1995; Belloni et al. 1996). 

Further studies in timing reveals more differences among the
states from the extensive observations of Cyg X-1 with Ginga and
RXTE, such as, the various shapes of power density spectrum (PDS) (see, e.g., van
der Klis 1995; Belloni et al. 1996; Cui et al. 1997), various spectrum evolutions
in both short time scale ($\sim$1 second) (Feng, Li, \& Chen 1999; Li, Feng, \& Chen 1999)
and long time scales ($\sim$several hours) (Wen, Cui, \& Bradt 2001). 
Although there is no model which provides a consistent explanation of both the temporal and
the spectral properties of the states, the various states have been attributed to different
geometric structures and physical properties of the accretion flow (e.g., Esin, McClintock
\& Narayan 1997) which may be closely related to the accretion rate and determined
by the configuration of the mass transfer (e.g., Gies et al. 2003).

Spectroscopy of Cyg X-1 in the various states can greatly enhance
the knowledge of the configuration of the mass transfer in the system.
Red-shifted absorption lines of highly ionized species have been
detected with HETG/Chandra during the low state at orbital phase 0.73 (Schulz et al. 2002), 0.84
(Marshall et al. 2001) and  0.77 (Miller et al. 2002a) and in intermediate states
at orbital phase 0.88 (Feng et al. 2003 in preparation).
These lines show that there is a highly ionized inflow with a variable configuration.
This inflow may play an important role in controlling the spectral state and in triggering
the state transition.
However, the previous observations of Cyg X-1 with Chandra covered only a limited range of
orbital phase (0.73-0.88) and spectral states (the low and intermediate state) which prevents us
from obtaining a complete picture of the stellar wind.
Spectroscopy of Cyg X-1 in both the high and the low states around orbital phase 0
is particularly interesting because the column density may be enhanced and produce strong absorption
lines.
Furthermore, comparison of the properties of absorption lines detected in the different
states and orbital phases will show the variations in the configuration of mass density
 and shed light on the mysteries of the spectral state and its transition.

The fluorescent Fe K$\alpha$ line may be a powerful way to probe the accretion disk in
Cyg X-1 (Fabian et al. 1989).  Fe K$\alpha$ line emission detected during the low state
indicates that there is a cold accretion disk where the iron fluorescent
emission line is generated (e.g., Barr, White and Page 1985; Ebisawa et al. 1996;
Miller et al. 2002b).
The detection of a $\sim$6.6 keV broad line emission in 1996 during the
intermediate state with ASCA (Cui et al. 1998) has been attributed to Fe K$_\alpha$
emission line from the highly ionized inner region of the accretion disk rather than
the cold disk of the low state.
Additionally, in the high state, the inner edge of the accretion disk may extend down to
the last stable orbit. The high resolution obtained with HETG/Chandra in the detection of the
possible Fe K$_{\alpha}$ line emission will constrain the line emission model.
In this paper, we present results from the first spectroscopy of Cyg X-1 in the high
soft state around orbital phase 0 with HETG/Chandra.

\section{Observation and Analysis}
Monitoring of Cyg X-1 with ASM/RXTE shows that from Sept. 2001 to Oct. 2002
the X-ray flux was 2-4 times higher than $\sim$20 cts/sec of the low state,
as seen in figure 1, indicating that the source might have entered the soft state
(Zhang et al. 1997). Furthermore, the hardness ratio $<$ 0.6 and positive
correlation coefficient between hardness ratio and total count rate were found
during the observation occurred, as seen in in figure 1. These are consistent
with that Cyg X-1 had entered the high soft state (Wen et al. 2001).
A Chandra X-ray Center Director's Discretionary Time (DDT)
observation (ObsID: 3724) proposed to catch Cyg X-1 in the high state
occurred from July 30, 2002 17:27:17 to July 31, 2002, 01:52:08
with a simultaneous RXTE observation.
PCA/RXTE spectrum of Cyg X-1 was found to be the typical shape of the high state and power
density spectrum was in the shape of power-law. All these timing and spectral properties confirmed
that Cyg X-1 was in the genuine high soft state during the observation.
Also the companion star was closest along our line of sight which is the so
called orbital phase 0 based on the ephemeris provided by Gies et al. (2003).

(Figure 1)

The HETG/Chandra observation operated was in continuous-clocking (CC) graded mode in order
to eliminate pile-up. To reduce telemetry saturation, a 100 pixel region centered on the
zeroth order image was transmitted with a 10\% duty cycle. However, because of the
very high X-ray  flux, the observation still suffered from telemetry saturation
and frame-dropping.

In CC mode ACIS reads out the CCD in virtual frames.  This ensures
that all the on board processing is identical for CC and TE modes.
Due to the fact the CCD is clocked differently there is a slight gain
difference between these modes.  This gain is only used for order
sorting, and we have verified the order sorting is still selecting
the correct data.  For high count rate sources, data are generated
faster than the telemetry can handle.  If ACIS has processed a frame
of data, which represents about 1.5 sec of CC mode data from one CCD
and that data will not fit in the telemetry stream, it discards the
entire frame.  This results in a data gap for that CCD but otherwise
has no impact on the data or processing.  However, it is important
to realize that each CCD will now have a different integration time,
which is easily calculated from the ``Good Time Intervals" given in
the FITS file.

To extract the spectrum of a point source from the grating instrument of
HETG/Chandra, it is important to precisely locate the source position on the CCD.
We located Cyg X-1's position on ACIS-S CCD chip S3 in three steps.
First, we located the Cyg X-1 position from its zeroth order image.
Then, we compared it with the location calculated from the galactic coordinate
of Cyg X-1. Finally, we adjusted the source position to make the spectral features
in both minus and plus order spectra line up. Based on the source
position, we used CIAO 2.2.1 and followed the updated threads provided
by CXC to extract the spectrum. The HETG/Chandra spectrum was also extracted with
a software developed by us. The spectral features reported here are consistent
in both spectra. For a bright X-ray source, such as Cyg X-1, the background
detected with HETG/Chandra in 0.7-4 keV region is negligible. The spectrum in this range
is analyzed without background subtraction. When we searched for the broad feature near 6.5 keV,
we did subtract a background which contributed about 4 percent of the counts in this band.
 The background was estimated at a given location from ACIS-S
energies just above and below the dispersed grating energy for the location.

Before looking for the broad Fe K$_{\alpha}$ emission line, we examined
the HEG spectrum in the range of 4 to 5 keV. It showed that the plus
and minus orders do not agree. The data which fall
near the edge of S3 are about 10 percent lower than the corresponding
data from S2.  At a distance about 250 pixels from the S3 edge the
two rates roughly agree.  This ripple can also be seen in the data
from the ACIS calibration source (see Fig. 5 of Townsley et al. 2002)
and is weak at small chipy but increases as chipy increases.  This
ripple is due to a combination of serial and parallel clocking charge
transfer inefficiencies.

To correct for this ripple effect, we applied a simple correction
factor to our ARF files.  From the Mn line calibration source data,
at the chipy coordinate where the Cyg X-1 data were taken, we know
that the efficiency is about 10 percent lower near the readout and close
to full efficiency at the maximum distance (256 pixels).  For the
model we assumed that the correction was linear between these two
points.  This correction was applied to all 4 nodes on S3 resulting
in a zig-zag correction.  When we used the corrected ARF we now find
much better agreement between the plus and minus orders.

When we jointly fit Chandra and RXTE spectra in the range between
2.5 to 15 keV, we allowed the RXTE absolute
effective area to vary by a constant (independent of energy) amount.
It is worth noting that although the RXTE and Chandra data overlap
in time, the fact that the Chandra data were continuous, whereas the
RXTE data were broken by gaps, means that the data are not from
precisely the same time interval. Thus some of the differences could
be due to the source variability.  Furthermore we added 1.0 percent systematic
uncertainty to the RXTE data.

\section{Results}
Fitting the spectrum with a single power-law in 2-10 keV energy range gives a power-law index
about 3.2, and although it was not a good fit, it does indicate that Cyg X-1 was in the high
state during the observation. A better fit can be obtained by adding either a second power-law
or a disk black body. In the residuals of the fitting absorption lines of highly ionized
species show up significantly. A broad feature near 6.5 keV
is also detected. In this paper we report the broad and asymmetric absorption lines
and the broad emission feature. Comprehensive studies of the absorption lines will be
reported elsewhere.

(Figure 2)

\subsection{Asymmetric Absorption Lines}

Strong and asymmetric absorption lines are detected in Cyg X-1 during the
high state. The line profile is  composed of an extended red wing and a less
extended blue wing, as seen in figure 2.
The red wings of higher ionized species, such as, Mg XII, Si XIV, S XVI, and Fe XXII,
are more extended than that of lower ionized species, such as, Mg XI, Si XIII, S XV, and Fe XIX,
as shown in figure 2 and table 1.
Comparison of the absorption line of Mg XII detected in the different orbital phases and states
shows that the bottom of the absorption lines are near zero velocity (see figure 3), whereas
those of previous observations in the low and intermediate state are red-shifted.
It shows also that an absorption line of the high state has a more extended blue wing
than that of the low and intermediate states.
During the high state (phase 0) the red wing is more extended than
that in the low state (phase 0.74) and that in the intermediate state (phase 0.88),
as shown in figure 3.

\begin{table}
\footnotesize
\begin{center}
\caption{Selected Absorption Lines Identification and Properties}
%\vskip 4pt
\begin{tabular}{lllcccccc}\hline
Ion& Rest & Meas.$^a$& Shift & Red Wing$^b$ & Blue Wing$^b$ & EQW & Nz$^c$ & R$^d$ \\
 & ($\AA$)&($\AA$) &(km/s) & (km/s) & (km/s)&(m$\AA$) & (10$^{15}$cm$^{-2}$) & \\
\hline
Mg XII &8.4192 & 8.4252$^{+0.0018}_{-0.0014}$
 &213$_{-63}^{+48}$
 & 1776
 &-534
 &10.0(0.3) & 57.7(1.6) & 2.6(0.2) \\
Mg XI & 9.1688 & 9.171$^{+0.014}_{-0.005}$ 
 & 55$_{-466}^{+171}$ 
 & 1794
 & -655
 & 7.5(0.4) &13.6(0.6) & 4.9(1.2) \\
Si XIV & 6.1822 & 6.1857$^{+0.0005}_{ -0.0005}$
 & 167$_{-25}^{+22}$
 & 1935
 & -850
 & 10.8(0.4) &77.0(2.8) & 3.9(0.3) \\
Si XIII & 6.6480 & 6.6469$^{+0.0004}_{ -0.0004}$ 
 & -48$_{-19}^{+19}$ 
 & 1350
 & -903
 & 6.0(0.4) &20.4(1.5) & 5.4(0.9) \\
S XVI & 4.7229 & 4.7323$^{+0.0018}_{-0.0004}$
 & 593$_{-115}^{+25}$
 & 1264
 & -688
 & 6.6(0.6) & 80.1(7.0) & 6.0(1.5) \\
S XV & 5.0387 & 5.0401$^{+0.0010}_{-0.0015}$
 & 80$_{-58}^{+92}$
 & 1039
 & -1043
 & 4.2(0.5) &24.0(3.0) & 4.4(0.4)\\
Fe XXII & 11.770 & 11.7711$^{+0.0013}_{-0.0011}$ 
 & 28$_{-34}^{+28}$
 & 1905
 & -701
 & 19.3(0.7) &23.2(0.8) & 2.8(0.2) \\
Fe XIX &13.518 & 13.5134$^{+0.0006}_{-0.0013}$
 & -102$_{-13}^{+29}$ 
 & 499
 & -499
 & 10.2(0.8) &16.4(1.3) & 3.0(0.5) \\
 \hline 
 \end{tabular}
 \end{center}
$^a$ The wavelengths are calculated by fitting Gaussians to the line features.
 1$\sigma$ errors are quoted.\\
 $^b$ Width is distance between the edge of wing and the deepest absorption, as seen in figure 2.\\
 $^c$ The column densities of the absorption lines are calculated by using:
 $W_{\lambda} = \frac{\pi e^{2}}{m_{e} c^2} N_{j} \lambda^{2} f_{ij}, $\\
\noindent where $N_{j}$ is the column density of a given species,
$f_{ij}$ is the oscillator strength, $W_{\lambda}$ is the equivalent
width of the line which is calculated by fitting the spectrum in a narrow 
range to a powerlaw and a negative Gaussian, and $\lambda$ is the wavelength of the line in cm
units (Spitzer 1978). \\
 $^d$ Ratio between the column densities derived from HEG spectra observed at phase 0 and at phase 0.77. 
\normalsize
\end{table}

(Figure 3)

\subsection{Broad Emission Feature}

We first fit the joint RXTE and Chandra spectra to a simple continuum model consisting
of two power-laws.  This gave a $\chi^2$ of 1573.6 for 835 degree of freedom (dof)
and the residuals are plotted in the top panel of figure 4a.  We next added
three Gaussian, two with negative norms (at 2.62 and 6.67 keV) for
the hydrogen-like Si and the He-like Fe lines.  The third broad Gaussian
was added at about 6.5 keV. The new $\chi^2$ was 1060.8 for 826 dof and the broad line has
a central energy of 6.49$\pm$0.04 keV and width $\sigma$ of 0.51$\pm$0.04 keV.
If we replace the low energy power-law with a disk blackbody component
then $\chi^2$ becomes 1028.4, but the line is now centered at 5.44$\pm$0.12 keV
with width $\sigma$ 1.42$\pm$0.07 keV. Errors are 68\% confidence range for a single paramenter.

We tried to replace the broad Gaussian with the more physical Laor              
(1991) component.                                                               
At no time did the Laor component generate a better $\chi^2$ than the             
Gaussian and often $\chi^2$ was significantly worse.                              
Thus with the two power-law continuum, the Laor resulted in $\chi^2$ = 1061.7     
and for this case the best fitting parameters for the disk had an inner radius  
of 25 r$_g$ (r$_g$=GM/c$^2$), an outer radius of 100 r$_g$ and an inclination of 86$^o$.                 
With the disk black body plus power-law continuum, the best $\chi^2$                
obtained was 1266.4 for a "disk" extended from 23.5 to 26 r$_g$ and                 
inclination of 86$^o$.                                                          
For both these models we included the two negative gaussians for the            
Si and Fe absorption lines.                                                     
We also note that often the fitting routine would make the negative             
Fe line broad and positive, effectively discarding Laor and formally,           
this did result in a better fit.                                                
Physically reasonable Laor models tend to be asymmetric and the                 
fitting tended to find non-physical solutions which made the                    
line symmetric.    
Although an asymmetry could be smeared by the lower
resolution of the RXTE spectrum, the higher resolution of the HETG
makes it impossible to hide such an asymmetry.  Part of the problem is due
to the disk blackbody and power-law components being roughly equally important
in the 5-6 keV range.  Since the disk blackbody spectrum is falling rapidly,
the precise nature of the fall is sensitive to assumptions
made. Merloni, Fabian and Ross (2000) have shown that self consistent disk models
produce more flux at higher energies than the XSPEC diskbb model predicts. 
Thus the excess seen in 5-7 keV may be partially attributed to the under estimation
of the flux at higher energies with the diskbb model.

(Figure 4)

\section{Discussion}
Strong absorption lines of highly ionized species are detected in Cyg X-1 during
the high state around orbital phase 0, as seen in figure 3.
This suggests that there is a high mass density in the region between the companion
and the black hole, as shown in the simulation of high mass binaries (Blondin et al. 1991).
Similar species of absorption lines are found during observations in the high, the intermediate,
and the low states.
The slight change of the ionization state of stellar wind is consistent with the X-ray flux
increase by only a factor of 2-4 in the high state from that in the low state,
if the photoionization is the dominated ionization process in Cyg X-1.
Strong absorption lines of Fe XXV to Fe XVII have been detected, indicating a
large range of the ionization state in the wind.
The large variation in the ionization state is not surprising since
the density of the wind decreases as it moves away from the companion
and the irradiation flux depends on the distance to the compact object.
This combination results in a large variation in the ionization
parameter along the line of sight.

The absorption lines are broad and asymmetric, and consist of an extended red
wing and a less extended blue wing.
The extended red wing can be attributed to the inflow with a projected radial velocity
larger than that of out-flow.
The inflow is the region where the angle between the motion direction of the material
and our line of sight is larger than 90${^o}$.
According to the wind structure of the radiatively driven wind model of Castor, Abbott, \& Klein (1975),
this region is closer to the X-ray source, than the region where the blue wing is generated.
The closer to the source the wind is, the higher photoionization parameter
and higher ionization degree it should have.
Indeed, the red wings of higher ionized species, such as, Si XIV, S XVI, and Fe XXIII,
are found to be more extended than that of lower ionized species, such as, Si XII, S
XV and Fe XIX, as seen in figure 2 and table 1, consistent with the above picture of mass transfer.
As the wind material approaches the X-ray source, its projected radial velocity increases
because of the strong gravity around the black hole; in addition
the ionization parameter should also increase because the wind receives a higher X-ray flux.
This picture is consistent with the existence of the lower (or non) ionized stellar
wind at high-latitude (far from the source) where the blueshifted absorption H$_{\alpha}$
line is produced and has been detected by Gies et al. (2003).
The projected radial velocity of blueshifted absorption line of H$_{\alpha}$ is
$\sim$ 380 km/s which is comparable with that of the blue wing seen in figure 2.
The asymmetric distribution of the projected radial velocity implies a focused stellar wind
toward the black hole.

The 5.6 day orbital modulation was detected in the X-ray emission of Cyg X-1 during the low
state (Holt et al. 1979; Brocksopp et al. 1999; Wen et al. 1999) which is consistent
with the distribution of X-ray intensity dips measured with ASM/RXTE
(Ba{\l}uci{\'n}ska-Church et al. 1999) and PCA/RXTE (Feng \& Cui 2002). It suggests enhanced
mass density in the region where the companion faces the black hole during the low state.
However, no orbital modulation was detected from X-ray flux (2-12 keV) during the high
state (Wen et al. 1999). 
Since the velocity distribution of the stellar wind with high ionization degree
has been found to be broader than those in the other states, as seen in figure 3.
We think that one possible reason for the lack of orbital modulation is that the stellar wind
is highly ionized in a larger region during the high state. Therefore, the column density of cold
stellar wind could be lower and produce weaker modulation during the high state which is consistent
with the speculation of Wen et al. (1999). 
Another possibility is that the distribution of the stellar wind in the region where X-ray emission has been
attenuated becomes more isotropic from the view point of observer during the high state.
If this is true, the difference between the stellar wind in the low and the high states implies that
state transitions are triggered by variations in the configuration of the stellar wind.
Comparison with a low state HETG observation near phase 0 will help resolve which model is more correct.

Part of the highly ionized stellar wind in Cyg X-1 flows away from the black hole and
the other part is accreted by the black hole.
There are two possible ways for the accreted material to fall into the horizon of the black hole, i.e.,
through the corona or the accretion disk.
If it is via the disk, the highly ionized stellar wind needs to condense
to form the cold disk, for example, via comptonization cooling of low-energy photons
(Igumenshchev et al. 1999). As the highly ionized plasma is cooled down to form the cold disk
recombination emission lines are expected.
However, no recombination line emission has been detected.
Therefore, the highly (or fully) ionized stellar wind most likely goes into the horizon mainly
via the corona. The existence of the corona during the high state is consistent with the
presence of the hard/non-thermal component in Cyg X-1 during the high state, as well as in
the low and intermediate states.
In studies of the X-ray intensity dips, two types of dips have been found (Feng \& Cui 2002).
One type (called type A) exhibits strong energy-dependent attenuation
of X-ray emission at low energies during a dip, which is characteristic
of photoelectric absorption from cold material. The other type (called type B) shows nearly
energy-independent attenuation which is consistent with the partial obstruction of the
X-ray emission region by dense streams.
The dense stream could be cold and neutral which may exist mainly around the
equatorial plane and form the disk.

Comparison of the absorption lines observed in different orbital phases shows that
the blue wing is more extended at phase 0 than those at
 other orbital phases. The blue wings may be attributed to the stellar wind flowing away
 from the source. Less extended blue wings and red shifted absorption lines suggest
a focused wind structure.
However, the variation may also be attributed to the change of the spectral state,
because the absorption lines have been detected not only at different orbital phases but
also in different states.

Based on the radiatively driven wind model (Castor, Abbott, \& Klein 1975),
the column density viewed from phase 0 is expected to be about two times higher than that
viewed from phase 0.77. The column densities in the lines observed at phase 0 in the high state are
 more than two times of those at phase 0.77, as seen in table 1. This implies stronger
stellar wind in the high state which can cause higher accretion rate. However, this
interpretation is dependent upon the wind model which may not suitable to apply to Cyg X-1.
Again comparison with HETG/Chandra observation of Cyg X-1
around orbital phase 0 in the low state will reveal the difference in stellar wind between
the low and high states. Spectroscopy of Cyg X-1 with more complete orbital
phase coverage in the low and high states is necessary to probe the structure of the mass transfer
between the companion and the black hole in Cyg X-1.

There is a very broad and significant feature in the spectrum near
6.5 keV that can be modeled with a broad Gaussian function.  Attempts to
model the feature with a Laor function have resulted in worse fits.
This is most likely due to the Laor function being asymmetric whereas
the residuals seen in figure 4a appear fairly symmetric.  The precise
shape of this excess depends on the continuum model, with small
changes in the continuum resulting in large changes to the excess.
Thus we have been unable to use this broad feature to derive physically
meaningful parameters. Nevertheless, the appearance of this broad feature
in the high state reveals that the spectral evolution in Cyg X-1 is much
more complicated than that detected with low resolution detectors.
The spectral evolution suggests that the geometric structure and physical
properties of the accretion flow may be associated with the spectral state.

\acknowledgments
We thank Chandra Director Harvey Tananbaum, and all the other members of the
Chandra team for their enormous efforts to conduct the DDT observation.
We thank all the  members of the RXTE team for their enormous efforts to conduct
a simultaneous observation with RXTE. We thank helpful discussion with
M.C. Weiffkopf, M.K. Joy and K.K. Ghosh. We gratefully acknowledge the support for
this work was provided by NASA through Chandra Award Number GO2-3061X and DD2-3018X
issued by Chandra X-ray Observatory Center, which is operated by Smithsonian
Astrophysical Observatory for and on behalf of NASA under contract NAS8-39073.
This work was also supported in part by NASA Marshall Space Flight Center under
contract NCC8-200 and by NASA Long Term Space Astrophysics Program under grants
NAG5-7927 and NAG5-8523.

\clearpage

\begin{figure}
\plotone{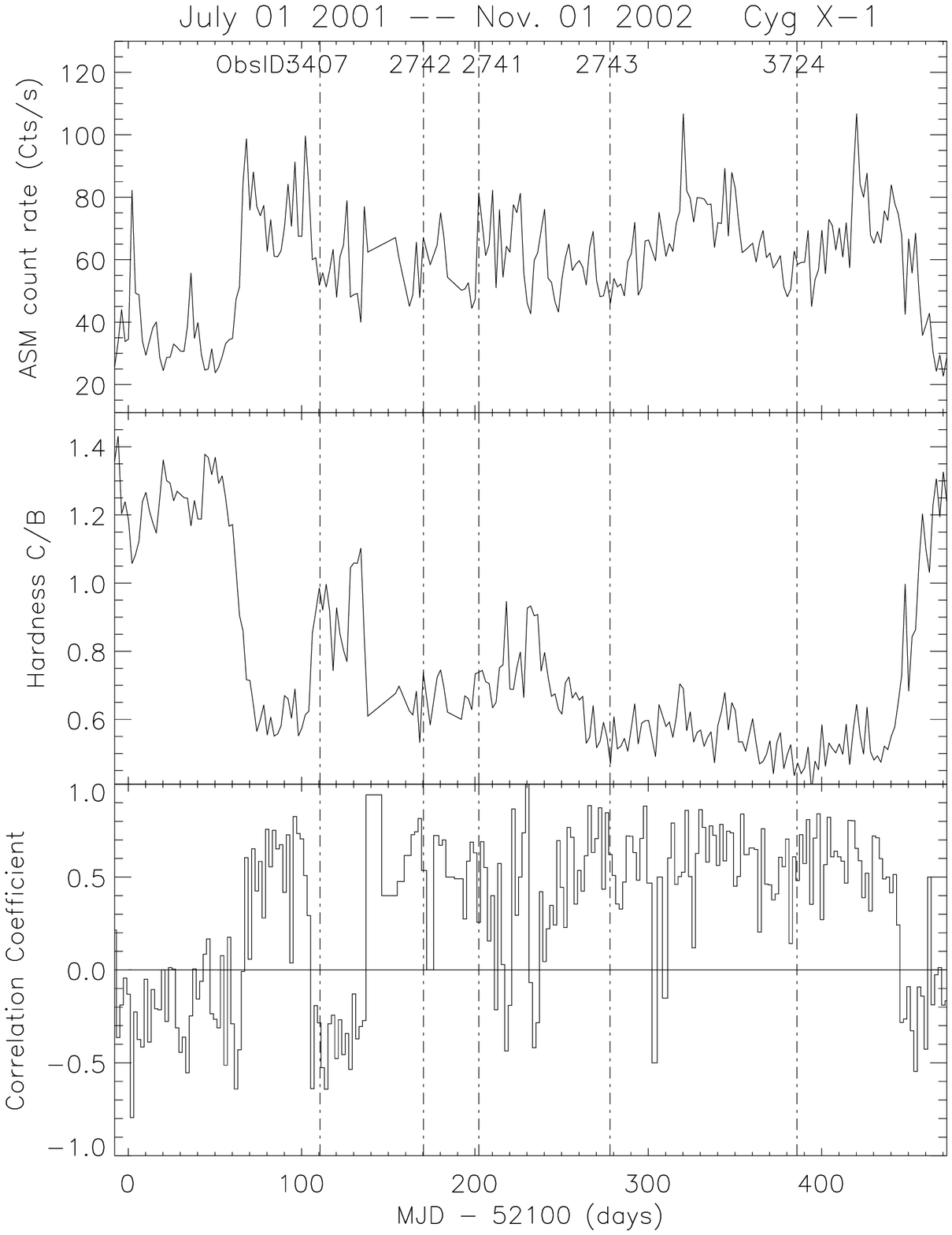}
\caption{The ASM/RXTE monitoring of Cyg X-1 from July 1, 2001 to Nov. 01, 2002.
The vertical lines in the plot shows observation time with HETG/Chandra. }
\end{figure}

\clearpage
\begin{figure}
\plotone{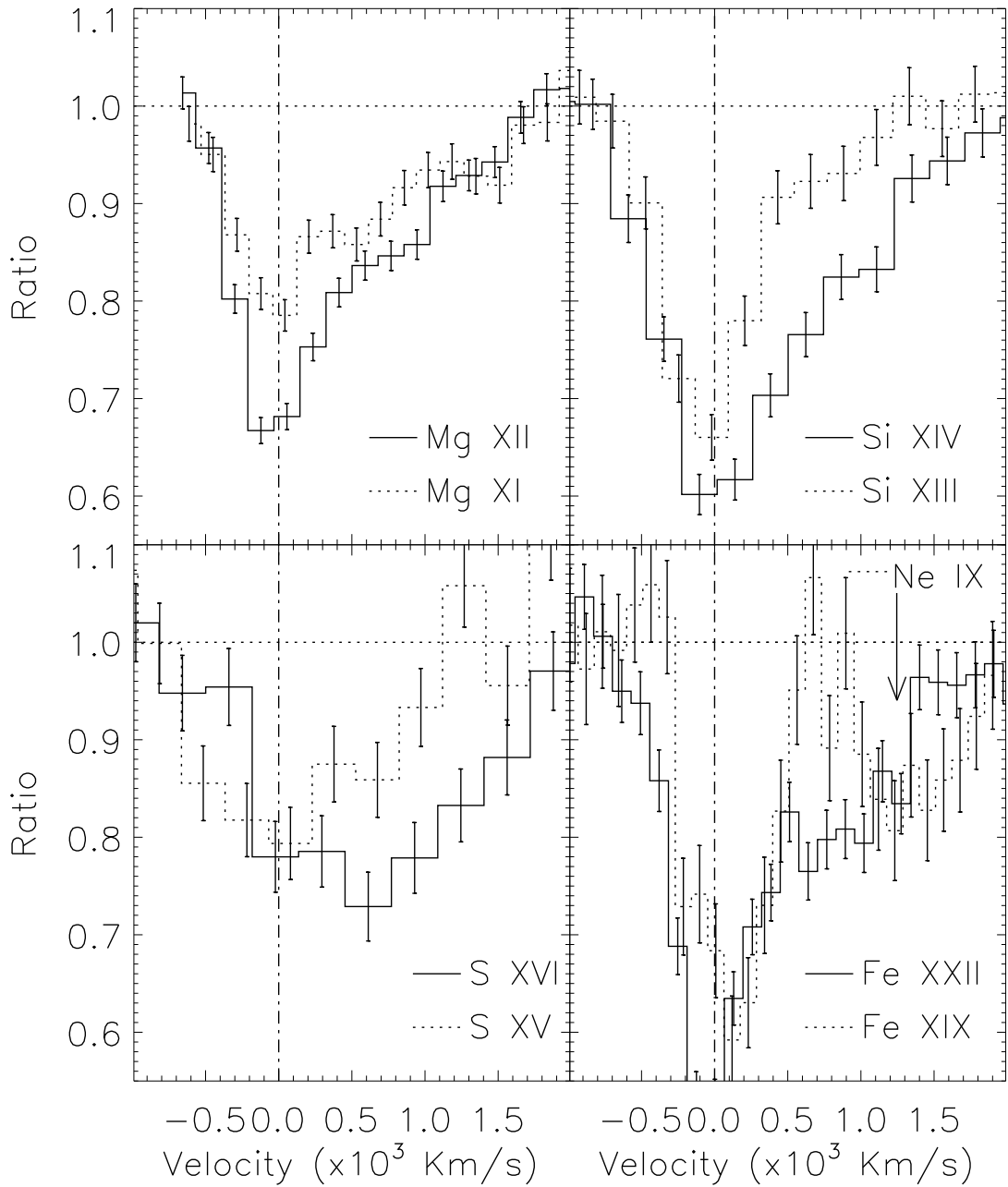}
\caption{The strong absorption line features were observed in the high state July 30, 2002
around orbital phase 0. The red wing of lower ionized species is less
extended than that of higher ionized species. }
\end{figure}

\clearpage
\begin{figure}
\plotone{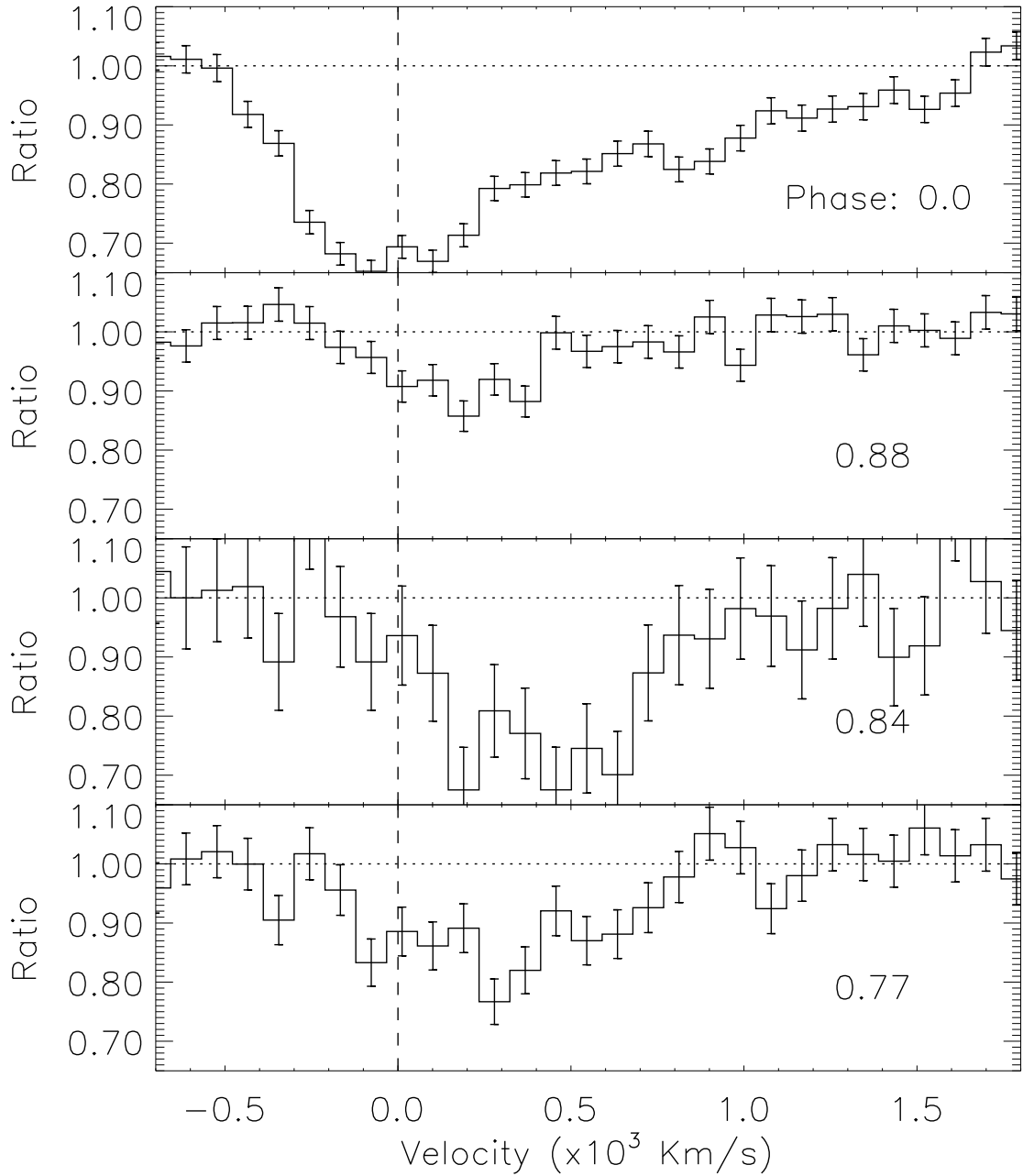}
\caption{The comparison of Mg XII absorption line (8.4192 $\AA$) features
observed in the high state July 30, 2002 around orbital phase 0 with that in the low
state, observed on Jan. 04, 2001 at the phase 0.77 and Jan. 12, 2000 at the phase 0.84, and
that in the intermediate state observed in Oct. 28, 2001 at the phase 0.88. }
\end{figure}

\clearpage
\begin{figure}
\rotatebox{-90}{
\epsscale{0.8}
\plotone{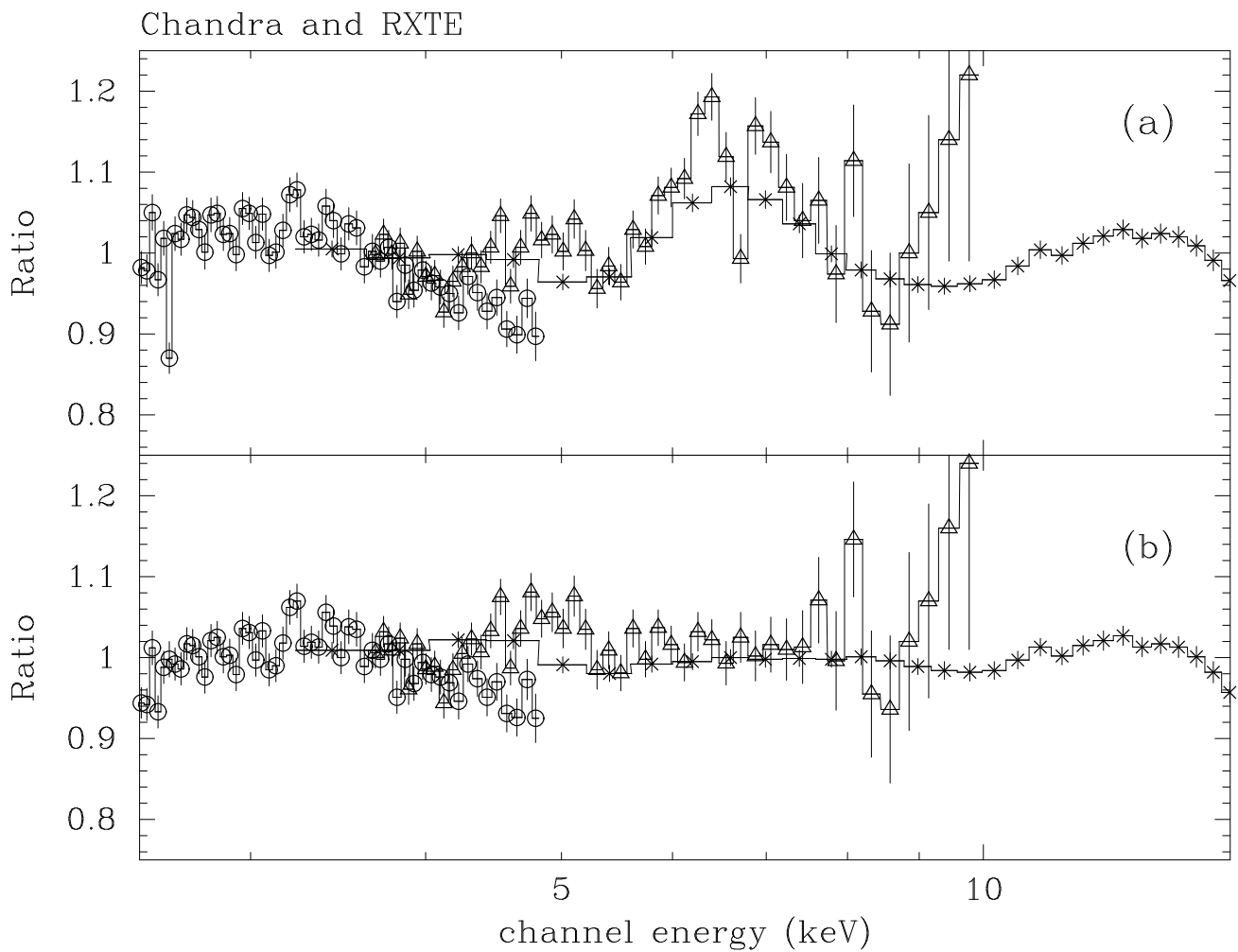}}
\caption{The broad emission feature is detected from jointly
 fitting spectra of Cyg X-1 during the high state observed with
 HETG/Chandra and PCA/RXTE. a) the residuals of fit with
 a simple continuum model consisting of two power-laws; b) the residuals of fit
with the same continuum model as that of a) and three Gaussians, two with negative
norms (at 2.62 and 6.67 keV) for the hydrogen-like Si and the He-like
Fe lines and one broad Gaussian at about 6.50 keV.}
\end{figure}

\end{document}